\documentclass[
    ,final            % use final for the camera ready runs
  ]
  {aipproc}

\layoutstyle{6x9}

\newcommand{\gtrsim}{\mathrel{\hbox{\rlap{\lower.55ex \hbox {$\sim$}}
                   \kern-.3em \raise.4ex \hbox{$>$}}}}
\newcommand{\lesssim}{\mathrel{\hbox{\rlap{\lower.55ex \hbox {$\sim$}}
                   \kern-.3em \raise.4ex \hbox{$<$}}}}

\begin{document}

\title{The INTEGRAL Galactic Bulge monitoring program}


\author{E. Kuulkers}{
  address={ISOC, ESAC/ESA, Apartado 50727, 28080 Madrid, Spain}
}

\author{S.E. Shaw}{
  address={University of Southampton, UK}
  ,altaddress={ISDC, Switzerland}
}

\author{S. Brandt}{
  address={DNSC, Denmark}
}

\author{J. Chenevez}{
  address={DNSC, Denmark}
}

\author{T.J.-L. Courvoisier}{
  address={ISDC, Switzerland}
}

\author{K. Ebisawa}{
  address={ISAS, Japan}
}

\author{P. Kretschmar}{
  address={ISOC, ESAC/ESA, Apartado 50727, 28080 Madrid, Spain}
}

\author{C.B. Markwardt}{
  address={University of Maryland, USA}
  ,altaddress={NASA/GSFC, USA}
}

\author{N. Mowlavi}{
  address={ISDC, Switzerland}
}

\author{T. Oosterbroek}{
  address={ESA-ESTEC, The Netherlands}
}

\author{A. Orr}{
  address={ESA-ESTEC, The Netherlands}
}

\author{A. Paizis}{
  address={INAF-IASF, Italy}
}

\author{C. Sanchez-Fernandez}{
  address={ISOC, ESAC/ESA, Apartado 50727, 28080 Madrid, Spain}
}

\author{R. Wijnands}{
  address={University of Amsterdam, The Netherlands}
}

\begin{abstract}
The Galactic Bulge region is a rich host of variable high-energy point sources.
These sources include bright and relatively faint X-ray transients, X-ray bursters, persistent neutron star
and black-hole candidate binaries, X-ray pulsars, etc.. We have
a program to monitor the Galactic Bulge region regularly and frequently with
the $\gamma$-ray observatory {\em INTEGRAL}. As a service to the scientific
community the high-energy light curves of all the active sources as well as 
images of the region are made available through the WWW. We show the first 
results of this exciting new program.
\end{abstract}

\maketitle

%%%%%%%%%%%%%%%%%%%%%%%%%%%%%%%%%%%%%%%%%%%%
%% MAINMATTER
%%%%%%%%%%%%%%%%%%%%%%%%%%%%%%%%%%%%%%%%%%%%

\section{Introduction}

The bulge of our Galaxy hosts a variety of
X-ray and $\gamma$-ray point sources
(e.g., Knight et al.\ 1985, Skinner et al.\ 1987, Churazov et al.\ 1994;
see, e.g., Bird et al.\ 2006, B\'elanger et al.\ 2006, Revnivtsev et al.\ 2004, 
for {\em INTEGRAL} observations).
These include persistent and transient neutron star and
black-hole candidate binaries, X-ray pulsars, X-ray bursters, etc.. 
Because of the variability these sources possess on time scales of msec to days
(quasi-periodic oscillations, pulsations, [absorption] dips, eclipses, type I and
type II X-ray bursts, orbital variations, flares) and weeks to years (orbital
variations, outburst cycles, on/off states), the region never looks exactly the same.

From 17 February 2005 onwards {\em INTEGRAL} has been monitoring this region
approximately every 3 days, as part of our approved AO-3 program, whenever the region
is visible by {\em INTEGRAL}.
In this paper we describe this program in more detail and show the first results of the
first two periods of monitoring performed between February and October 2005.

\section{INTEGRAL and data analysis}

{\em INTEGRAL} (The {\bf Inte}rnational {\bf G}amma-{\bf R}ay {\bf A}strophysics {\bf L}aboratory; 
Winkler et al.\ 2003) is an ESA scientific mission dedicated to fine spectroscopy 
($E/\Delta E$$\simeq$500; SPI) and fine imaging (angular resolution: 12 arcmin FWHM; IBIS)
of celestial $\gamma$-ray sources in the energy range 15\,keV to 10\,MeV
with simultaneous monitoring in the X-ray (3--35\,keV; JEM-X) and optical
(V-band, 550\,nm; OMC) energy ranges.

Our program is to
observe the region frequently and regularly,
with the aim to investigate
the source variability and transient activity on time scales of days to weeks to months
at relatively soft ($\lesssim$10\,keV) and hard ($\gtrsim$10\,keV)
energies.
One complete hexagonal dither pattern (7 pointings of 1800\,sec each, i.e.,
1 on-axis pointing, 6 off-source pointings in a hexagonal pattern around
the nominal target location, each 2$^{\circ}$ apart)
is performed during each {\em INTEGRAL} revolution, or orbit around the earth (i.e., roughly every 3 days).
This is done whenever the region is visible by {\em INTEGRAL}
(about two times per year for a total period of about 4 months).
As a service to the scientific community, the JEM-X light curves (3--10\,keV
and 10--25\,keV) and the IBIS/ISGRI light
curves (20--60\,keV and 60--150\,keV) are made
publicly available as soon as possible after the
observations are performed. In addition, IBIS/
ISGRI and JEM-X mosaic images of each hexagonal
observation are provided, with information on
the detected sources. Last, but not least, all
IBIS/ISGRI 20--60\,keV mosaic images per revolution
are stacked into a movie, showing the
ever-changing gamma-ray sky.
All the instruments onboard {\em INTEGRAL}, except the OMC, have coded masks.
With the fully and partially coded field of views (FOVs) we cover about half of the 
low-mass X-ray binary (LMXB) and high mass X-ray binary (HMXB) Galactic Bulge 
population (see also, e.g., in 't Zand 2001).

Similar Galactic Bulge monitoring programs have been
performed (see, e.g., in 't Zand 2001) and are currently ongoing
(e.g., {\em RXTE} Galactic Bulge Scans, Swank \&\ Markwardt 2001).
However, the {\em RXTE}/PCA and HEXTE do only have a 2$^{\circ}$ collimator, so only
a small field of view with no imaging
resolution, and therefore only provide information on a given source for a short time
when the instrument scans over it; moreover, in the Galactic Center region
itself there is some source confusion. There are currently other instruments in operation 
at similar energy ranges (e.g., {\em Swift}/BAT: 15--150\,keV with a FOV of 2 steradians; Barthelmy 2000), but 
they do not provide frequent and regular monitoring of the Galactic Bulge region, 
as well as having a worse imaging capability, again leading to some source confusion 
in the Galactic Center region ({\em Swift}/BAT PSF angular resolution is 22' compared to the 
IBIS/ISGRI angular resolution of 12' [FWHM]).

For our program at the moment we only consider data from IBIS/ISGRI (Ubertini et al.\ 2003,
Lebrun et al.\ 2003) and JEM-X (Lund et al.\ 2003). 
We do not consider the data from the
IBIS/PICsIT, SPI, or OMC instruments. 
Either the angular resolution is high
(SPI: 2.5$^{\circ}$) and therefore the various sources
in the Galactic Bulge region close to each other complicate the
analysis, or the sources are too weak to be detected (IBIS/PICsIT).
For the OMC, however, we are currently evaluating the scientific output.

The {\em INTEGRAL} data reduction is performed using
the Off-line Scientific Analysis ({\tt OSA};
Courvoisier et al.\ 2003), v5.1. We use a source catalog,
currently containing 79 sources which have been detected by IBIS/ISGRI up to now
in the field we are interested in (but see next Section). 

The data from IBIS/ISGRI are processed
until the production of images in the 20--60 and 60--150\,keV energy ranges
per single exposure. We force the flux extraction of each of the 
catalogue sources, regardless of the detection significance of the source.
This method is essential in 
order to clean the images from the ghosts of all the active sources in 
the field, but does not make any threshold selection and all the positive 
fluxes are recorded.
In order to detect fainter sources, we then mosaic the images from the single exposures and 
search for all catalog sources, as well as possible new ones.
For JEM-X the analysis is run through the imaging step to the light-curve step in {\tt OSA}
for a single bin of the same length as the exposure window. Light curves
are produced for all catalog sources up to 5$^{\circ}$ off from
the center of the FOV. Again, the images from the single exposures
are mosaiced in order to create the final image (but no further 
source detection was done).

Per hexagonal dither 
(i.e., 7 exposures combined) we are sensitive down to typically
between 5 and 15\,mCrab (6$\sigma$) for both JEM-X and IBIS/ISGRI. 
The actual sensitivity depends on factors such as source position 
(fully or partially coded FOV), background (instrument systematics, 
solar activity), number of exposures (some are lost) and energy (instrument response).

The results, as well as more information about
the program, can be retrieved from the {\em INTEGRAL} Galactic Bulge Monitoring WWW
page hosted at the ISDC in Switzerland: \url{http://isdc.unige.ch/Science/BULGE/}.

\section{Galactic Bulge monitoring: first results}

By now we have had two full seasons of monitoring,
i.e., from revolutions 287--307 (2005 February - April) and 347--370
(2005 August - October), respectively.
The third season started in revolution 406 (February 2006).
In Figure 1 we show examples of results from the first two seasons
for various types of sources.
At the left we show the light curves of (temporary) bright 
(i.e., easily detected in one exposure) sources. At the right 
we show the light curves of weaker (persistent or transient) 
or slowly varying sources, averaged per revolution. 

\begin{figure}
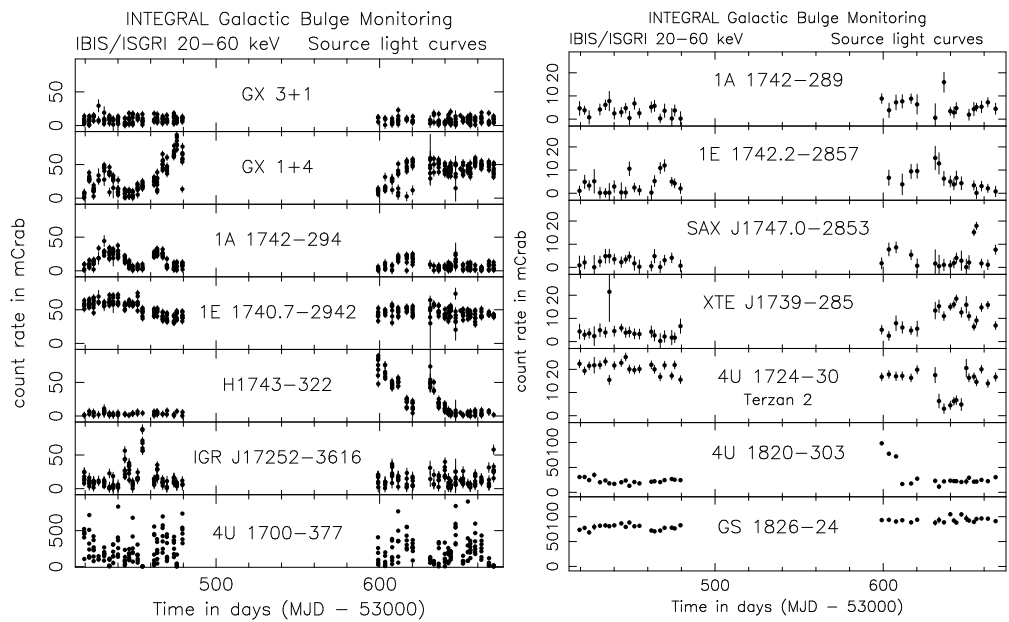

  \includegraphics[height=.3\textheight,angle=-90]{lc_20-60_osa51_scw.ps}
  \includegraphics[height=.3\textheight,angle=-90]{lc_20-60_osa51_rev.ps}
  \caption{{\em INTEGRAL} IBIS/ISGRI (20--60\,keV) light curves from the first two
seasons of the Galactic Bulge monitoring program. Shown
are some examples of the sources detected. 
The light curves are either on a time scale of one exposure, i.e., 1800\,sec
({\it left}) or on a time scale of one revolution, i.e., roughly 3 days
({\it right}).}
\end{figure}

Whereas, for example, GX\,3+1 (neutron star LMXB) does not vary much, 
sources like GX\,1+4 (symbiotic binary containing
a neutron star) and 1A\,1742$-$294 (neutron star LMXB) vary
on monthly times scales, while 1E\,1740.7$-$2942 (LMXB) varies smoothly
on even longer time scales. Some sources clearly show transient
behaviour, i.e., they show outbursts with durations exceeding
months (e.g., H1743$-$322, a black-hole candidate LMXB; see also below) or
flaring on timescales of hours to days (e.g., IGR\,J17252$-$3616,
an X-ray pulsar HMXB).
Some sources vary on all timescales accessible through our program,
as displayed by the HMXB 4U\,1700$-$377.

Similarly, sources like 1A\,1742$-$289 and 1E\,1742.2$-$2857
(both unidentified X-ray sources) show low-level activity
on various timescales.\footnote{These 
sources are not
included in our source catalog, but the light curves displayed here
are the result when the ISDC Reference catalog is used as input. Note
that these sources were not reported by B\'elanger et al.\ 2006 and
Bird et al.\ 2006, based on long exposure times. Further investigation
is in process.}
Sources like the neutron star LMXB transients SAX\,J1747.0$-$2853
and XTE\,J1739$-$285 showed renewed outburst activity, during the second season
(see also below). The neutron star LMXBs 4U\,1724$-$30 (in the
globular cluster Terzan~2) and 4U\,1820$-$303 have been persistently
on through the seasons, displaying occasionally drops or flares, 
respectively, in intensity for about a month. The neutron star
LMXB GS\,1826$-$24 slowly varied through our observing periods.

So far quick-look results during the two seasons have been reported in
10 ATel's. Here we describe some of the highlights.
Precisely at the start of the program the black-hole
X-ray transient GRO\,J1655$-$40 was reported to
become active (Markwardt \&\ Swank 2005). 
The {\em INTEGRAL} GRO\,J1655$-$40 light curves (see Kuulkers et al.\ 2005a)
nicely complement observations at soft
X-ray ({\em RXTE}; see Homan 2005) and radio ({\em VLA}; see Rupen et al.\ 2005) 
wavelengths. Various other transient sources popped up and
faded away, such as The Rapid Burster, H1743$-$322
(both Kretschmar et al.\ 2005; for H1743$-$322 
see Figure 1, left), IGR\,J17098$-$3628 (Mowlavi et al.\ 2005), 
SAX\,J1747.0$-$2853 (Kuulkers et al.\ 2005b; see Figure 1, right)
and XTE\,J1818$-$245 (Shaw et al.\ 2005a).
In 2005 August, the X-ray transient XTE\,J1739$-$285 
was found by {\em INTEGRAL} to be bright at soft and not detected at
hard X-ray energies (Bodghee et al.\ 2005). About a month
later the situation had reversed; it was bright at hard
and weak at soft X-ray energies (Shaw et al.\ 2005b; see Figure 1, right).
Although at first we attributed the state change to
the compact object being a black hole, we proved it 
to be a neutron star based on the occurrence
of type I X-ray bursts detected with JEM-X (Brandt et al.\ 2005).

\section{Conclusions}

We have shown that most of the sources in the program 
in the field of view of the {\em INTEGRAL} instruments 
clearly vary on timescales of a few
hours to days to months; it is therefore of no surprise that the
Galactic Bulge is a region to stay tuned
on. MIRAX with its wide-field instruments covering
a similar energy range (Braga et al.\ 2004; see also these
Proceedings) will go a step further, i.e., it will 
{\it continuously} monitor the Galactic Bulge region 
for about 9 months per year down to a sensitivity level of $\simeq$5\,mCrab
per day. Our monitoring program is, therefore, also an ideal 
`training session' for what to expect with MIRAX.

%%%%%%%%%%%%%%%%%%%%%%%%%%%%%%%%%%%%%%%%%%%%%%%%
%% BACKMATTER
%%%%%%%%%%%%%%%%%%%%%%%%%%%%%%%%%%%%%%%%%%%%%%%%

\begin{theacknowledgments}
Based on observations with {\em INTEGRAL}, an ESA project with instruments and science data centre funded 
by ESA member states (especially the PI countries: Denmark, France, Germany, Italy, Switzerland, Spain), 
Czech Republic and Poland, and with the participation of Russia and the USA. 
\end{theacknowledgments}

\bibliographystyle{aipproc}   % if natbib is available

\end{document}